# Differential hydrophobicity drives self-assembly in Huntington's disease

Martin G. Burke, Rüdiger Woscholski[†] and S.N. Yaliraki *

Department of Chemistry, [†] Department of Biological Sciences,

Imperial College London, South Kensington Campus, London, SW7 2AZ, UK

*Corresponding author: S.N. Yaliraki

Department of Chemistry

Imperial College London

South Kensington Campus

London, SW7 2AZ

U.K.

s.yaliraki@imperial.ac.uk

tel: +44 207 594 5899

fax: + 44 207 5945880

Classification: *Biological Sciences*: Biophysics

Text pages: 24

Figures: 4

Abstract Words: 216

Total number of characters 39,646 (with spaces)

Abbreviations:

        Poly-Q  (poly-glutamine)

        DPD (Dissipative Particle Dynamics)

**Identifying the driving forces and the mechanism of association of huntingtin-exon1, a close marker for the progress of Huntington's disease, is an important prerequisite towards finding potential drug targets, and ultimately a cure. We introduce here a modelling framework based on a key analogy of the physico-chemical properties of the exon1 fragment to block copolymers. We use a systematic mesoscale methodology, based on Dissipative Particle Dynamics, which is capable of overcoming kinetic barriers, thus capturing the dynamics of significantly larger systems over longer times than considered before. Our results reveal that the relative hydrophobicity of the poly-glutamine block as compared to the rest of the (proline-based) exon1 fragment, ignored to date, constitutes a major factor in the initiation of the self-assembly process. We find that the assembly is governed by both the concentration of exon1 and the length of the poly-glutamine stretch, with a low length threshold for association even at the lowest volume fractions we considered. Moreover, this self-association occurs irrespective of whether the glutamine stretch is in random coil or hairpin configuration, leading to spherical or cylindrical assemblies, respectively. We discuss the implications of these results for reinterpretation of existing research within this context, including that the routes towards aggregation of exon1 may be distinct to those of the widely studied homopolymeric poly-glutamine peptides.**

Neurodegenerative disorders are often linked with insoluble protein aggregates of fibrillar morphology, rich in β-structure content. In Huntington's disease, aggregates of N-terminal proteolytic fragments (exon1) of the protein huntingtin (1, 2) are found in the nuclei or the peri-nuclear cytoplasm of neurons (3, 4). Although a major thrust of research (5) is focused on the pathogenic role of huntingtin exon1 association, the underlying driving forces and mechanism of this process, which could ultimately provide a therapeutic approach towards overcoming HD, remain to be established (6).

The age of onset of HD is correlated with the expansion of the CAG trinucleotide repeat sequence which encodes for glutamine, with a pathogenic threshold of 34-41 consecutive glutamines (poly-Q) (7). Because of this observation, previous research has predominantly focussed on the propensity of long homopolymeric poly-Qs to form hairpin or other *β*-sheet structures as a prerequisite and driving force for the formation of insoluble fibrillar aggregates (8). Perutz's influential proposal (7) — that hydrogen bonding between the main chain and side chain amides could lead to stabilised polar zipper structures only for poly-Q lengths exceeding the threshold — shares among current models the emphasis on the length-dependent random coil to *β*-sheet structure transition of the single poly-Q peptide chain.

However, recent experiments have demonstrated instead that poly-Q in solution is in a stable random coil conformation irrespective of its length (9). Further studies have corroborated this finding for exon1 fragments over a broad length range of the poly-Q stretch (10-13). Above a concentration threshold, exon1 aggregates form *in* *vitro* in a concentration dependent process (14), and only after prolonged existence as pre-fibrillar globular-type suspensions. Given that the poly-Q component at the pathogenic threshold forms less than half of the peptide fragment (see Fig. 1) it is surprising that only recently have biophysical studies addressed the aggregation properties of entire exon1 fragments.

The exon1 fragment sequence consists of a block of glutamines followed by a stretch of mainly proline amino acids, an arrangement that remarkably resembles diblock copolymers, where two homopolymeric blocks are covalently linked. (Fig. 1) (15). This key realization, particularly since proline is chemically very dissimilar to glutamine in water, coupled with the above experimental findings, leads us to draw an analogy to this class of systems. It is well known that block copolymers in solution will spontaneously self-assemble into complex mesoscopic morphologies (16). Their shape (17), size (18), stability and dynamical behaviour (19), as well as possible geometrical transformations (20), intricately depend on a number of factors (16-20) whose relative contributions are still unfolding. These are: (i) the effective interaction between monomers of each block, (ii) each block's affinity for the solvent, (iii) the length of each block and (iv) their relative volume fraction (where conformation plays a role). A key implication for our system is that in a *selective* solvent (that is, solvent that interacts differently with each block), spontaneous self-assembly into spherical or cylindrical structures is expected in dilute solutions, with the more hydrophobic part forming a core and the more hydrophylic part an outside corona. The critical condition for occurrence of this directed self-assembly depends on the concentration as well as the length of each block . To date, little attention has been paid to the implications of the block structure of exon1 and the difference in hydrophobicity of glutamine and proline residues as a driving force for the initiation of exon1 association. This is the focus of this paper.

To explore the mechanism of self-association of the exon1 fragment theoretically as a function of concentration and chain length, it is crucial that the method can treat a large number of chains over long timescales without getting trapped in kinetic barriers. The solvent has both to be explicitly taken into account and also exhibit the correct hydrodynamic behaviour. This makes simulations extremely challenging: all-atom, or even coarse-grained, Molecular Dynamics are prohibitive, Monte-Carlo are less insightful and still difficult to equilibrate (21); and Brownian dynamics have also been shown to get trapped in metastable configurations (22). To overcome these limitations, we have used a recent off-lattice particle-based methodology, Dissipative Particle Dynamics (DPD) (23) which has been shown to correctly lead to mesoscale structures in block copolymer melts (24) and cell membranes (25, 26). This is, to our knowledge, the first application of this method to peptides in solution. Simulations in this work are based on effective particles (beads) at the residue level. The solvent is explicitly modelled and preserves hydrodynamics. The relevant interactions are obtained through a systematic procedure based on a map to Flory-Huggins theory (27). As a result, the DPD method enables us to study the dynamics of systems at least 3 orders of magnitude bigger than previous protein aggregation studies (28).

Equipped with these tools, our simulations reveal that spontaneous association of the diblock exon1 fragments occurs when the length of the poly-Q segment, $N_Q$, is as low as 18 at volume fractions as low as 2.5%. Assemblies readily form *irrespective* of whether the poly-Q segment is initially in a random coil or hairpin configuration (Fig. 1b), with the assemblies being sphere- or cylinder-like, respectively, without any *a priori* assumption on shape. This occurs without any explicit attractive interactions in the model. The implication is, not only that β structure is not a necessary condition for association, but, also, the initial assembly route towards the insoluble fibrils may not be available to the poly-Q-only peptides.

**Methodology**

**DPD fundamentals**. The DPD method was originally proposed (23) to describe the hydrodynamics of atomic fluids. The effective radial forces acting on the unit particles, or, *beads*, are pairwise additive, short-ranged, and have no hard core. All forces are zero beyond a cut-off distance $R_c$, which defines the only lengthscale in the system, and hence the size of the beads for a given bead number density ρ. Taking the bead mass $m$ as the unit of mass, the unit of time $\tau$ is then given by $\tau = R_c\sqrt{m/kT}$. The total force on each bead $i$ is given by a conservative, dissipative and random component $f_i = \sum_{j \neq i}(\mathbf{F}_{ij}^C + \mathbf{F}_{ij}^D + \mathbf{F}_{ij}^R)$, where the sum runs over all beads $j$ within radial distance $R_c$. The forces act along the bead centres and, importantly, conserve linear and angular momentum. The dissipative forces arise from the lost internal degrees of freedom of the bead, and are linear in relative velocity, while the random forces arise from their coupling to the environment. It has been subsequently shown that if the magnitudes of the dissipative $\mathbf{F}^D = -\gamma\omega^D(\hat{\mathbf{r}}_{ij} \bullet \mathbf{v}_{ij})\hat{\mathbf{r}}_{ij}$, and random $\mathbf{F}_{ij}^R = \sigma\omega^R\theta_{ij}\Delta t^{-1/2}\hat{\mathbf{r}}_{ij}$ components satisfy the fluctuation-dissipation theorem, the simulation obeys the canonical distribution and a constant temperature can be maintained (29). Following

Groot & Warren (27), the arbitrary weight functions, $\omega^D$ and $\omega^R$, are $\omega^D(r) = [\omega^R(r)]^2 = (1-r)^2$ for $r < R_c$ and 0 for $r \geq R_c$. The noise amplitude ($\sigma$) and drag constant ($\gamma$) are related by $\sigma^2 = 2\gamma k_B T$. $\mathbf{r}_{ij}$ is the distance between the centres of beads $i$ and $j$, $\mathbf{v}_{ij}$ their relative velocity and $\hat{\mathbf{r}}_{ij}$ the unit vector joining their centres. $\theta_{ij}(t)$ is a randomly fluctuating variable with Gaussian statistics. The conservative part of the force can encode physico-chemical properties to the beads and is given by $\mathbf{F}_{ij}^C = a_{ij}(1 - r_{ij}/R_c)\hat{\mathbf{r}}_{ij} + \mathbf{F}_{polymer}$ for $r_{ij} < R_c$ and 0 otherwise, where $a_{ij}$, the maximum repulsion between beads $i$ and $j$, is obtained through an involved but systematic procedure (25, 27), and $\mathbf{F}_{polymer}$ describes the appropriate spring forces that create polymers from beads (27).

**Modelling**

**Map for coarse-graining the conservative force**.
Although the accuracy of interatomic interactions is sacrificed, and indeed meaningless at lengthscales smaller than the bead size, beads still retain physico-chemical properties through the magnitude of the repulsive conservative force parameters, $a_{ij}$, which have two components, $a_{ij} = a + [\Delta a]_{ij}$. For a given bead size (and hence bead density $\rho$), the magnitude of the repulsion parameter $a$, common to all beads, can be derived from the equation of state of the system in order to match the compressibility of the solvent (27), since we are in the dilute solution regime. In our system using this approach self-consistently, we obtained a repulsion parameter $a$ =239 for density $\rho$=5, and hence a bead size of 450 Å$^3$, which corresponds to approximately 3 residues. This result is further justified when considering the conformations of each of the blocks (see **Conformations**). To model mixtures, the excess repulsion parameters for the interactions of unlike beads $[\Delta a]_{ij}$ are obtained through a procedure (25, 27) by analogy to and in quantitative agreement with Flory-Huggins solution theory of immiscible polymers. The free energy of mixing within this theory is given in terms of the phenomenological parameter $\chi$ which accounts for the interactions between species (30). Following Refs. (25, 27), a relationship between $\chi$ and excess repulsion can be obtained, which for our system, was found to be *$\chi=0.63\ \Delta a$*. We note that this procedure remains valid only within the range where mean-field is expected to hold or homogeneous mixing can be numerically achieved. This opens the way to further explore relating microscopic information to mesoscopic lengthscales.

**Quantifying $\chi$.** A distinct value of $\chi$ is required to model each pair of interactions in our system within Flory-Huggins theory, shown also to be applicable in biomolecular systems (31). Although $\chi$ can be measured by light scattering or partition experiments, we could not identify such experiments in the literature either for polyproline or for polyglutamine in an aqueous solution. The relevance of these measurements motivate this experimental work. Deriving $\chi$ values from microscopic considerations for realistic systems is not established at present and, is outside the scope of this work. We extracted the glutamine-proline interaction, $\chi_{Q-P} = -0.52$, from the effective inter-residue contact energies, which were obtained from averaging over crystal structures in protein banks with solvent molecules filling the voids (32). For the glutamine-water and proline-water interactions, we decided to focus on $\chi_{Q-W} = 2.26$ and $\chi_{P-W} = 0.39$ parameters as most suitable to reflect not only the enthalpic interactions (33) but also the capability of each monomer for hydrogen bonding with water, as well as its packing and conformation (34). It is important to note that taking only enthalpic considerations into account can lead to $\chi$ values which would imply that glutamine is considerably more hydrophilic than proline, while the solubility of each amino acid in water (154.5 g proline/100g water; 3.6 g glutamine/100g water), as an indicator of hydrophylicity, supports the opposite view. We have nevertheless explored an extensive series of other parameter combinations (see **Results**) and found our conclusions do not depend qualitatively on the absolute magnitudes of these parameters but mainly on their relative differences.

**Conformations.** The glutamine-water interactions described above underscore the importance of polymer conformation and its distinct behaviour from the monomer.

It is experimentally known that polyproline in aqueous solution adopts a relatively rigid $\alpha$-helical structure with 3 residues per turn, stabilised by a hydrogen bond between every fourth residue (35). This conformation, consistent with the restrictive motion of the pentagon loop, exhibits a persistence length of 220 Å (or 70 residues) which is an order of magnitude higher than other homo-polypeptides (typically 10-30 Å) (36). Thus polyproline behaves as a rigid rod, which is the shape we adopt here. It is important to note that the interdispersed residues between prolines in the exon1 polyproline fragment that we have ignored would make the chain semi-flexible at most and leave our conclusions and qualitative results unaffected (18). The rod-like helical conformation of the polyproline block was modelled by introducing an angle potential between consecutive proline beads.

The conformation of polyglutamine is still the subject of much debate in the literature. We use as the starting point the recent experimental evidence that the structure of poly-Q in monomeric form is random coil (13). We also consider pre-formed hairpin conformations as a contrast to examine the effects of conformational variation on the mechanism and kinetics of aggregation. All conformations are modelled through an additional force component, $\mathbf{F}_{polymer}$, to the conservative force. Individual beads are joined into a polymer chain by springs through the potential, $U_{spring}(i,j) = \frac{1}{2} k_{spring} (r_{ij} - r_{eq})^2$, where the subscripts $i$ and $j$ indicate connectivity in the chain ($j=i+1$ for linear chains or $j=\{i+1, N_Q-(i-1)\}$ for hairpins). The equilibrium bond distance, $r_{eq}=0.7$, and bond force constant, $k_{spring}=40.0$, are chosen such that the mean distance between connected beads equals the maximum of the pair correlation function of an equivalent system of unconnected beads (27). Angular harmonic potentials have been additionally used to model rod-like conformations for the polyproline block, with an equilibrium bond angle $\theta_{eq}=\pi$ and and an angle force constant, $k_{angle}=20$.

## Results

We performed simulations of exon1 fragments in concentrations ranging from 2.5 - 40% (2.5, 5, 10, 15, 20, 40) volume fraction in explicit water and for poly-Q segments covering the whole range of healthy to very pathogenic lengths, from 9 to 60 amino acids ($N_Q=9, 18, 27, 36, 60$). The proline blocks were in a rod-like conformation while the glutamine blocks were modelled as both random coils and $\beta$-sheet hairpins.

Simulations were carried out in two different box sizes, 20x10x22 and 20x20x22. Periodic boundary conditions were employed and in each case the bead number density (ρ) was set to 5 beads per unit volume. Since each bead represents approximately 3 residues or 15 water molecules, the DPD unit volume is equivalent to 2250 Å$^3$ and the unit length corresponds to 13.1 Å. Each polymer contained 9 proline beads, and for each of the different lengths of glutamine repeats 3 to 20 beads accordingly. For each length, we considered all volume fractions within the reported range. The box size was chosen to minimise finite size effects. All simulations were at constant temperature *T= 300K*.

To safeguard against biased clustering, the chains were always singly dispersed in solution at random positions with random initial (coil) configurations. Each run included at least 1x10$^5$ steps, with a maximum of 2x10$^5$. Due to the nature of the soft potentials in the DPD, correlations are lost faster than in other methodologies. In our calculations, we have averaged over configurations separated by 1000 timesteps and only after excluding typically the first 80000 steps. The leap-frog algorithm was used to propagate Newton's equations of motion thereby maintaining time reversibility (37) and a timestep of 0.02, checked for consistency within the stochastic differential equation, was used to maintain equilibrium within an 1% temperature range from approximately timestep 1000 onwards.

We have distilled the conclusions from our numerics in Figs. 2-4 and present the most representative rather than exhaustive results. Our results show that, at the concentrations considered, the poly-Q length onset for assembly is very low. Only for $N_Q$=9 did we observe chains remaining as monomers (Fig. 2a), in sharp contrast to all other cases where assemblies readily formed regardless of whether poly-Q was in a coil or hairpin initial conformation. A representative example where self-assembly occurs is shown in Fig. 2, for $N_Q$=36. The core of the assemblies is mainly formed by glutamine residues, with low water content and the proline blocks sticking out towards the water (in all figures prolines are in blue, glutamines in red). We consistently observed, for all lengths above the threshold, that random coil poly-Q segments formed sphere-like structures while hairpins led to cylindrical-like shapes. In both cases, assemblies did not fuse at low concentrations and chains did not readily leave the assemblies due to their high hydrophobicity. For the highest packing volume of 40% we observed the fusing of these assemblies.

Although visually apparent, we quantified the formation and shape progression of the clusters by calculating averages of a radial density distribution and a density profile along each of the three axes for the glutamine beads. Since the polyproline blocks do not form part of the core and their radial positions are largely determined by the last glutamine residue, they do not contribute additional information in describing the shape of these clusters. Using the final configuration of the simulation, and a specified bead centre-to-centre cutoff distance of 1.5, assemblies were defined according to standard procedure (38). Plots from the random coil simulations of glutamine bead density against radial distance from the centre of mass of the assemblies were then produced by averaging over equilibrated timesteps for each of the four assemblies formed for $N_Q$= 36 at 10% volume faction (Fig 3c, in green). The beads are clustered around the centres of mass of the assemblies with local volume fraction approaching 1 at the core of the assemblies demonstrating that there is little interpenetration of either proline or water beads at the centres of the assemblies. In contrast to the initial positions of the chains (in red), the peaks clearly show the spherical symmetry of the formed assemblies. Equivalent results are obtained for all structures formed from random coil poly-Qs.

Density profiles and isosurfaces were used to compare the structure of the assemblies formed when the glutamine blocks adopt random coil (Fig. 3a-b, 3d respectively) and hairpin conformations (Figs. 4a-b, 4d). Assemblies appear as peaks in the density profiles along the $x$ ( Fig. 3a) and $z$ axes (Fig. 3b) for the system shown also as an isosurface plot in Fig 3d. The four assemblies are highlighted by the two peaks along both axes. By contrast, in the hairpin simulation (Fig. 4 c) three elongated assemblies are formed. The density profile on the $x$-axis shows three clear peaks but the density fails to reach zero between the peaks. The anisotropic shape of the assemblies is highlighted by the difference in the density profiles along $x$ and $z$ axes (Fig. 4a, 4b). When the data is plotted as an isosurface (Fig. 4d) it appears that the elongated assemblies are in fact interconnected. These results hold for all formed assemblies from hairpin-containing exon1 fragments we considered.

The results we obtained were qualitatively robust for a broad series of additional parameter variations. We checked that the rod-like conformation of proline was a good approximation. We found clusters with inner cores of similar size and densities when the the angle potential $U_{angle}$ between proline beads was removed. The proline-glutamine strength interaction had a substantial effect only for packings of more than 40% volume, far from biological relevance. A similar effect was observed for the interaction strengths for poly-Q and polyproline with water. We checked this through exploring diverse interaction variations from $\chi_{GW}$ = - 0.34 to 2.26, and from $\chi_{PW}$ = 0.39 to 2.42. Finally, we note that even when we exchanged the hydrophobicity of the segments, the principles remained the same but the structures changed, namely the rigid structure of the prolines formed the core which led to cylindrical structures in both the random coil and hairpin poly-Q.

## Discussion

While there is a wealth of information concerning the pathology and physiology of Huntington's disease, the underlying forces and mechanisms that govern the aggregation of the exon1 fragment of huntingtin remain unclear. The similarity of exon1, composed primarily by a poly-Q block followed by proline-rich segments, to diblock copolymers, coupled with the well-established fact that block copolymers will spontaneously self-assemble into complex mesoscopic morphologies in selective solvent, motivated us to apply a mesoscale methodology capable of capturing the dynamics of exon1 fragments. This novel approach of using DPD for simulating protein aggregation established that: (i) the different hydrophobicities of the glutamine- versus the non-glutamine-segments are a major factor in the initiation of the assembly process; (ii) β-structure is not necessary for assembly to occur, although the processes of association for random coils and hairpins are distinct, (iii) the onset for this spontaneous association is governed by both the concentration and the length of the poly-Q stretch in the exon1 fragment.

The significance of differential or relative hydrophobicity for the initiation of assembly is that the physico-chemical behaviour of the exon1 fragment is not only dependent on the poly-Q properties, but, also, on the rest of the fragment. The difference in solvent affinity between the two blocks and its effect on the behaviour of the exon1 fragments in solution has important implications for the interpretations of existing and future experimental work, since investigations on exon1-fragments should show differences to that of pure poly-Q peptides. While the latter exclude the *relative* hydrophobicity properties by default, our framework may unify the various models presently discussed in the research community (5). For instance, our findings are in agreement with recent studies focussing on solubilising the normally insoluble poly-Q blocks by attaching soluble peptides or proteins, which reveal similar poly-Q length thresholds for the aggregation (3, 13, 14), and provide a microscopic explanation for the enhanced solubility. The notion of amphiphilicity-driven assembly of peptides has been put forward as an important concept for amyloid and PrP peptides, implicated in Alzheimer's and prion disease respectively (39). Our findings underscore this reasoning.

Quantifying the relative hydrophobicities of poly-Q and polyproline, the major amino acids present in the exon1 fragment is challenging. A $\chi$ value which reflects the hydrophobicity of the two blocks more accurately would also enhance the quantitative accuracy of our results. However, it is now recognised that $\chi$, which is a measure of solvent-solute interaction, includes entropic as well as enthalpic effects, and that it is composition, pressure and temperature dependent (33). Because the effective $\chi_{eff}$ varies as a function of conformation, it is clear that as a poly-Q segment transforms from a random coil to elongated $\beta$-sheet during the aggregation process, its packing characteristics will vary and $\chi_{eff}$ will therefore deviate from its apparent value in the globular state (34). It would hence appear that for poly-Q in aqueous solution $\chi_{eff}$ is a dynamically evolving parameter. Additionally, hydrogen bonding modulates all these effects. Not only does the change from random coil to a more elongated *ß-sheet* conformation tend to increase $\chi_{eff}$, but the reduction in available donor/acceptor sites for hydrogen bonding increases its value as well. This increase happens in a composition dependent manner and may have the effect of stabilising an aggregate further with increasing size. Thus, the hydrophobic character of poly-Q stretches may be stronger with increasing intra- and inter-molecular hydrogen-bonding (39).

The second major finding of this work is that assembly is possible *irrespective* of whether the poly-Q component is initially in random coil or hairpin configuration. The implication is that work focused on single-chain properties of poly-Qs may not be sufficient to describe all possible aggregation mechanisms of the disease. Although these investigations play an important role for characterising the initial monomer behaviour, the ensuing models allow only for the hairpin formation (or equivalent intra-molecular conformational changes to β-sheet containing structures) as a prerequisite for aggregation and neglect the role played by relative hydrophobicity in driving self-assembly. This is especially important because above the critical condition for assembly, the timescale for the formation of the self-assembled clusters may be much faster than any β-sheet structure formation within a single chain. Which of the two (or both) will occur depends on the timescale of such structured formations of the single chain exon1 fragments (which may itself be dependent on length). Recent experimental evidence suggests that the timescale for formation of the assemblies is in fact faster than that for the formation of β-content structures (14). From a theoretical point of view, the emerging picture is that, although the kinetics of random coils and hairpins are very distinct, the intermolecular and intramolecular degrees of freedom governing the thermodynamics as well as the kinetics of the system are coupled in non-trivial ways. For example, as the level of intra- and inter- molecular hydrogen bonding increases, this may in turn increase the hydrophobicity of the aggregates by lowering the number of hydrogen bond donor/acceptor groups available to interact with the surrounding water molecules. This increased hydrophobicity makes further inter- and intra-molecular hydrogen bonding more favourable due to the reduced local concentration of water molecules resulting in a positive feedback loop driving the system towards complete phase separation.

Another important outcome derived from our simulations concerns the onset for the spontaneous assembly. Applying our framework revealed that this process is favoured for poly-Q component lengths in the fragment as low as 18, and possibly shorter, which is indeed what investigators have observed, thus additionally validating our approach. Dynamic light scattering and NMR monitoring of the aggregation of peptides with glutamine stretches as low as 20 and 22 revealed that aggregation is possible for such short poly-Q segments (10, 13). It is also worth pointing out that our coarse-grained model contains *no explicit* attractive interactions among beads, so self-association is not an input but an outcome of the simulation. However, the lengths of the poly-Q are to be taken with caution, partly due to the fact that beads represent 3 residues, which results in an uncertainty of at least 3 Qs.

Finally, it is important to distinguish between the initial structures observed in our simulations, which may be stable or metastable, and the final precipitated or phase separated fibrils. Further steps towards the precipitates may include conformational changes *within* the core of the formed clusters, where the possibility of glutamine-based hydrogen bonding increases, and/or fusion between clusters. In either case, hydrogen bonding within and between glutamine stretches inside the cores must play an important role, and may eventually lead to Perutz's polar zippers and nanotube final fibrillar structures. Structural changes can now be interpreted within the context of preformed clusters and monitored initially within the solution. β–sheet content formation should not just be thought as occurring only in single chains, but also within many-chain clusters in solutions. Indeed, these soluble globular structures may have been inadvertently seen in Scherzinger et al.'s experiments (10) . Recently the work of Ross and co-workers has also emphasized the solubility of globular clusters (14). Using FTIR spectroscopy, they detected the appearance of progressively increasing secondary structure shortly after the formation of the globular oligomers. Crucially however a large increase in the band coincided with the appearance of fibres. This strongly supports the concept of oligomeric intermediates with limited hydrogen bonding acting as intermediate structures in the pathway to fibres with highly optimised hydrogen bonding networks or β-sheets.

In all the issues raised in the Discussion, the length of the poly-Q stretch plays a role because, as the more hydrophobic part of the exon1 fragment that readily forms the compact core of the clusters, it, rather then the hydrophilic block, mainly modulates all thermodynamic and kinetic interactions (40). Properties therefore become length-dependent in a progressive, rather in a switch-like manner. In conclusion, the finding that a contrasting hydrophobicity in stretches of amino acids initially induces spontaneous self-assembly should facilitate future research into extracting the complete aggregation pathways in HD as well as other poly-glutamine related diseases (11).

**Acknowledgements**

This work was supported in part by grants from the US Office of Naval Research, the EPSRC, and the BBSRC. M.G.B. thanks Sam Hughes for helpful discussions. We are grateful to Mauricio Barahona for a critical reading of the manuscript.

**Figure Captions**

**Fig. 1.** The huntingtin exon1 fragment is analogous to a block copolymer, where homopolymeric blocks are covalently linked in series (a). The hydrophobicity profile of the exon1 fragment containing a polyglutamine stretch (hydrophobic block, in red) and a polyproline stretch (more hydrophilic block, in blue) is sketched at the top of the figure. It provides a driving force for the spontaneous formation of self-assembled clusters. The initial 17 residues on the N-terminus side of the peptide were ignored since these are usually cleaved in the *in vitro* experiments (10). In the DPD simulations, the polyproline was modeled as a 'rod' due to its high persistence length, while the polyglutamine stretch was modeled in the two most prominent conformations: as a random coil or as a 'hairpin' (b). Each bead corresponds to approximately 3 residues. Simulations were run for $N_Q$ = 9, 18, 27, 36, 60 and for volume fractions 2.5, 5, 10, 15, 20, 40 % in water, modeled explicitly, for poly-Q in both random coil and rod-like conformations.

**Fig. 2.** Threshold for self-assembly is low and depends on concentration and length. Snapshots at timesteps (a) 1 and (b) $1x10^5$ for peptides with 9 and 36 glutamine residues at 10% concentration with the poly-Q block modeled as a random coil. Glutamine beads are in red, proline beads in blue and water beads are represented as points. 22000 beads were used for $1x10^5$ timesteps. The snapshots of part (b) are repeated without proline and water beads for clarity. No aggregation is obtained for $N_Q$=9, but for $N_Q$=36 a number of globular clusters are formed with the glutamine beads forming the inner core and the rod-like proline blocks sticking out. The proline blocks remain surrounded by water throughout the simulation demonstrating little

propensity to separate from water through clustering. Similar self-assembled structures are readily formed in all considered systems with glutamine stretches $N_Q$=18 and higher, for all volume fractions from as low as 2.5%. Note that initial configurations are always randomly chosen for dispersed chains in solution to avoid biased clustering. The methodology contains no explicit attractive interactions. Self-assembly is hence an outcome and not an input in the simulations. The procedures for deriving bead sizes and parameter interactions are outlined in the text. Here, we have used in DPD units: a=239, $a_{pw}$=239.61, $a_{gw}$=242.59, $a_{pg}$=238.17, σ=3, γ=4.5, T=1, timestep size = 0.02, ρ=5, in box dimensions 20x10x22.

**Fig. 3.** Assembly of exon1 fragments with poly-Q stretches in random coil conformation. The typical self-assembled formations for poly-Qs of any length above the threshold are globular with the glutamine beads dominating the core of the globules as shown by the isosurface plots (d). Proline beads distributed around the Q core and water beads are not included in the plots for clarity. The figure corresponds to poly-Q length $N_Q$=36 and 10% volume fraction. Comparing pre- and post-equilibration averaged density profiles in the x (a) and z (b) directions, shows the transition from a relatively uniform density distribution to an inhomogeneous distribution exhibiting significant local concentration fluctuations. The post-equilibrium volume fraction (c) approaches 1.0 at the core of each of the clusters, demonstrating the relative-hydrophobicity driven formation of the clusters.

**Fig. 4**. Assembly of exon1 fragments with poly-Q stretches in hairpin conformation. In contrast to Fig. 3, typical structures of poly-Q's with preformed hairpins are rod-

like (c). The post-equilibrium density profiles (a,b) highlight the anisotropy of the structures in contrast with those in Fig. 3. The isosurface plot (d) (which has been translated 12 $R_c$ along the x-axis for visualization purposes) shows that the rod-like clusters are in fact interconnected forming a single aggregate. Data shown is for same poly-Q length and volume fraction as in Fig. 3.

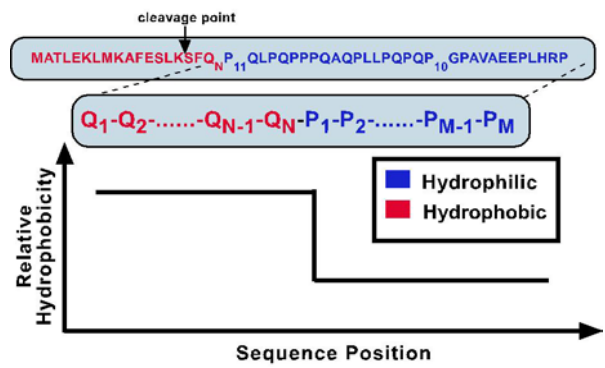
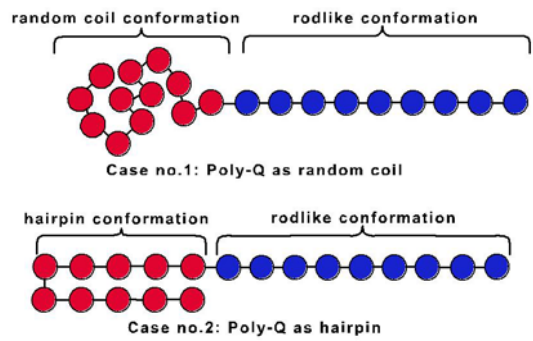

Fig. 1.

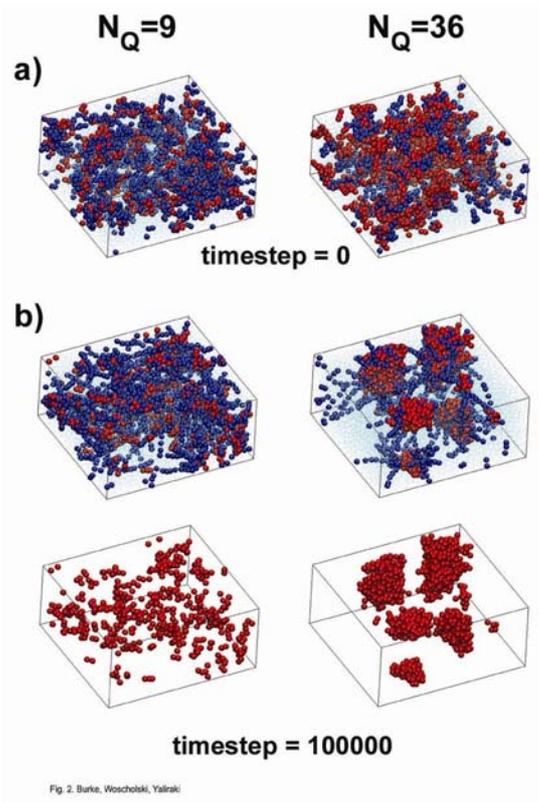

Fig. 2.

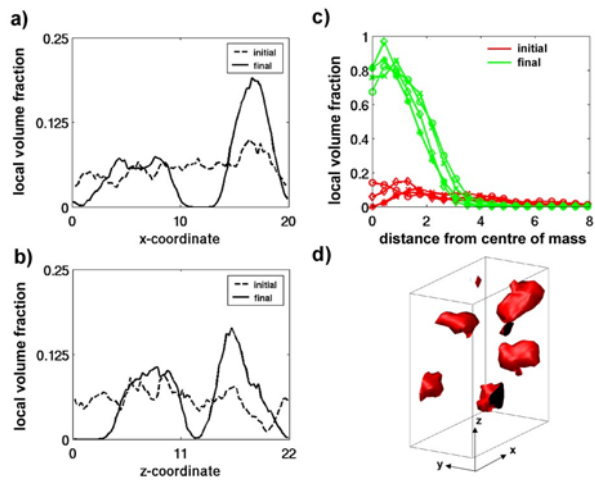

Fig. 3.

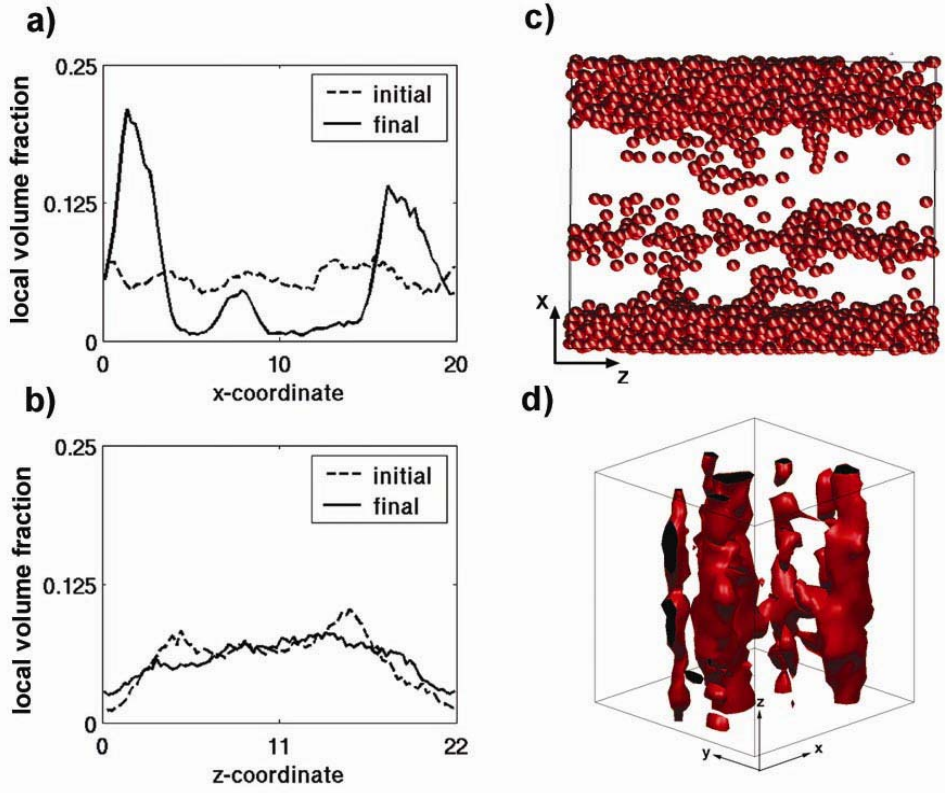

Fig. 4.